\documentclass[prb,twocolumn,showpacs,amssymb]{revtex4}

\usepackage{graphicx}
\usepackage{dcolumn}
\usepackage{bm}

\begin{document}


\title{Universality in the Energy Spectrum of Medium-Sized Quantum Dots}
\author{Alexander Odriazola$^a$, Alain Delgado$^b$, and Augusto Gonz\'alez$^a$}
\affiliation{$^a$ Instituto de Cibern\'etica, Matem\'atica
 y F\'isica, Calle E 309, Vedado, Ciudad Habana, Cuba\\
 $^b$ Centro de Aplicaciones Tecnol\'ogicas y Desarrollo Nuclear,
 Calle 30 No 502, Miramar,Ciudad Habana, C.P. 11300, Cuba}

\begin{abstract}
In a two-dimensional parabolic quantum dot charged with $N$ electrons,
Thomas-Fermi theory states that the ground-state energy satisfies the
following non-trivial relation: $E_{gs}/(\hbar\omega)\approx N^{3/2}
f_{gs}(N^{1/4}\beta)$, where the coupling constant, $\beta$, is the
ratio between Coulomb and oscillator ($\hbar\omega$) characteristic
energies, and $f_{gs}$ is a universal function. We perform extensive 
Configuration Interaction calculations 
in order to verify that the exact energies of relatively large quantum
dots approximately satisfy the above relation. In addition, we show that
the number of energy levels for intraband and interband (excitonic and 
biexcitonic) excitations of the dot follows a simple exponential 
dependence on the excitation energy, whose exponent, $1/\Theta$, 
satisfies also an approximate scaling relation {\it a la} Thomas-Fermi, 
$\Theta/(\hbar\omega)\approx N^{-\gamma} g(N^{1/4}\beta)$. We provide 
an analytic expression for $f_{gs}$, based on two-point Pad\'e 
approximants, and two-parameter fits for the $g$ functions.
\end{abstract}


\maketitle

\section{Introduction}

Thomas-Fermi theory \cite{TF,Kirzhnits,Lieb1,Spruch} has 
proven to be a valuable tool for the qualitative understanding of 
atoms and molecules. In semiconductor quantum dots \cite{KM,Hawrylak}, 
which are a kind of artificial Thomson atoms with many possibilities 
for fundamental research and technical applications, Thomas-Fermi 
theory was shown to agree qualitatively and even quantitatively with a
more ellaborated approach like Density Functional Theory 
\cite{Serra,Ullmo}, being asymptotically exact in the limit of large 
electron numbers \cite{Lieb2}.
  
From the computational point of view, Thomas-Fermi theory with minor
corrections is able to reproduce the ground-state energy of electrons 
in a quadratic potential \cite{TFmio} at the same level of accuracy of 
other semiclassical or semianalytic approaches like large-$D$ 
expansions \cite{PLA} or two-point Pad\'e approximants \cite{GPP}. 

In the present paper, we would like to stress on a less studied
aspect of Thomas-Fermi theory: the highly non-trivial scaling relations
following from it. We show that the number of electrons, $N$, and the
coupling constant, $\beta$, enter the ground-state energy in a scaled 
form. We perform extensive Configuration Interaction calculations for
quantum dots with $20\le N\le 90$ in order to verify this scaling. In
addition, on the basis of the numerical results, we show that similar
scaling relations are valid for the number of excited states in 
intraband and interband excitations. In this way, a universal 
parametrization of the density of energy levels in quantum dots is
provided.

We start with the Hamiltonian of a two-dimensional parabolic quantum 
dot charged with $N$ electrons. In oscillator units, the Hamiltonian 
can be written as:

\begin{equation}
\frac{H}{\hbar\omega}=\frac{1}{2}\sum_i \left\{p_i^2+r_i^2\right\}+
 \beta\sum_{i<j}\frac{1}{r_{ij}}.
 \label{eq1}
\end{equation}

The only approximations made in writting Eq. (\ref{eq1}) are the 
effective-mass description of electrons, the inclusion of an effective 
low-frequency dielectric constant, $\epsilon$, to model the medium, 
and the description of confinement by means of a harmonic-oscillator 
potential. These approximations are very common and well sustained
\cite{Hawrylak}. The coupling constant $\beta=E_{Coul}/(\hbar\omega)=
e^2 m^{1/2}/(4\pi\epsilon \omega^{1/2}\hbar^{3/2})$ is 
the ratio of Coulomb and harmonic-oscillator characteristic energies.

The fact that the number of electrons may enter the energy in a scaled 
combination with $\beta$ is, however, not trivial. Let us write the 
Thomas-Fermi energy functional \cite{Lieb2} for the present problem:

\begin{eqnarray}
\frac{E_{TF}}{\hbar\omega}&=&\int {\rm d^2}r
 \left\{\alpha n^2+n~r^2/2 \right\}\nonumber\\
 &+&\beta\int\int{\rm d^2}r{\rm d^2}r'\frac{n(r)n(r')}
 {|\vec r-\vec r'|}.
 \label{eq2}
\end{eqnarray}

\noindent
where $n(r)$ is the (surface) density at point $r$, and
$\alpha$ is a numerical constant. The above functional should be 
extremized under the constraint

\begin{equation}
N=\int{\rm d^2}r~n.
\label{constraint}
\end{equation}

Now, it is easy to realize that we can scale $r$ and $n$ in such a 
way that the l.h.s. of Eq. (\ref{constraint}) becomes one, and a factor 
$N^{3/2}$ is extracted from the r.h.s. of Eq. (\ref{eq2}). As a result, 
we get the following relation for the ground-state energy in the 
Thomas-Fermi approximation:

\begin{equation}
E_{gs}/(\hbar\omega)\approx N^{3/2} f_{gs}(N^{1/4}\beta).
\label{gsenergy}
\end{equation}

\noindent
Notice that the scaled Thomas-Fermi equations depend on a single
parameter, $z=N^{1/4}\beta$, which combines in a particular way the
coupling constant and the number of electrons.

\section{Scaling in the ground-state energy}

We first provide an analytical expression for $f_{gs}$ based on 
two-point Pad\'e approximants \cite{GPP} in the large-$N$ limit. It 
shows that the scaling predicted by Thomas-Fermi theory is quite 
general and compatible with true quantum effects.

Let us recall the definition of the $P_{4,3}$ Pad\'e approximant for 
the ground-state energy, given in Ref. \onlinecite{GPP}, which
interpolates between the $\beta\to 0$ (perturbation theory) and 
$\beta\to\infty$ (Wigner ``crystal'') expansions:

\begin{equation} 
P_{4,3}(\beta)=p_0+\frac{p_1\beta+
 p_2\beta^{2/3}(q_2\beta^{2/3}+q_3\beta)}
 {1+q_1\beta^{1/3}+q_2\beta^{2/3}+q_3\beta}.
\label{pade}
\end{equation}

\noindent
We use the large-$N$ asymptotic expressions for the coefficients 
\cite{GPP}, which lead to the following estimation for the ground-state
energy:

\begin{equation}
\frac{E_{gs}}{\hbar\omega N^{3/2}}\approx \frac{2}{3}+
 \frac{0.698~z+1.5~z^{4/3}+2.175~z^{5/3}}
 {1+2.149~z^{1/3}+1.5~z^{2/3}+2.175~z}.
\label{eq5}
\end{equation}

In order to verify the universal relation (\ref{eq5}) we performed 
extensive Configuration Interaction calculations for charged quantum 
dots. In these calculations, we follow standard procedures of
Quantum Chemistry \cite{CI} or Nuclear Physics \cite{Ring}.

\begin{figure}[t]
\begin{center}
\includegraphics[width=.95\linewidth,angle=0]{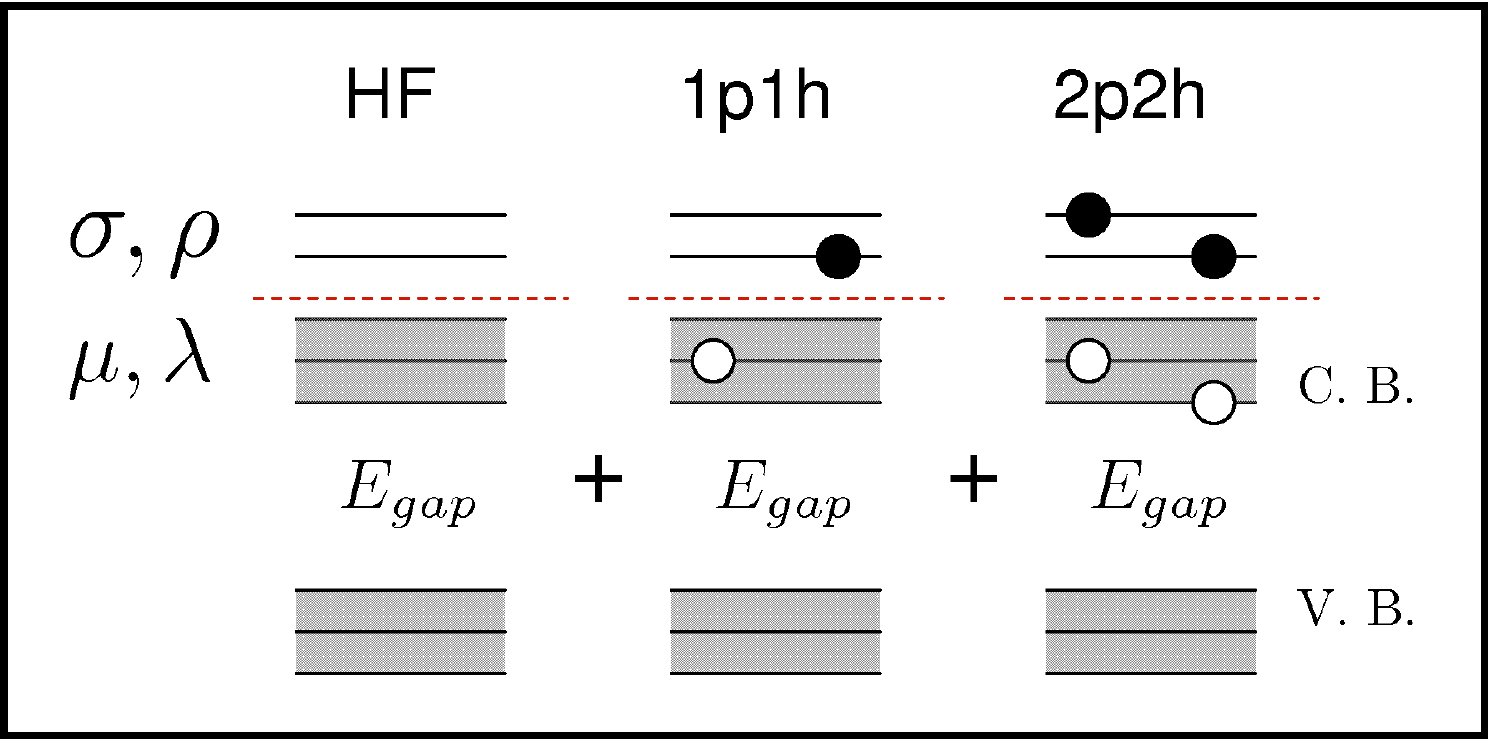}
\caption{\label{fig1} (Color online) Different contributions to the
ground-state wave function entering the Configuration Interaction 
calculation.}
\end{center}
\end{figure}

The starting point is the Hartree-Fock solution of 
the problem. Then a basis of functions made up from (i) 
the Hartree-Fock state, $|HF\rangle$, (ii) one-particle one-hole 
(1p1h) excitations, that is $|\sigma\mu\rangle=e^{\dagger}_{\sigma}
e_{\mu}|HF\rangle$, and (iii) two-particle two-hole (2p2h) excitations, i.e. 
$|\sigma\rho,\mu\lambda\rangle=e^{\dagger}_{\sigma}e^{\dagger}_{\rho}
e_{\mu}e_{\lambda}|HF\rangle$, is used in order to diagonalize the 
Hamiltonian. Notice that $\sigma<\rho$ are single-particle states 
above the Fermi level, and $\mu<\lambda$ are states below the Fermi 
level. A schematic representation is given in Fig. \ref{fig1}. In the
Hilbert subspace with the same quantum numbers of the Hartree-Fock
state, the electronic Hamiltonian takes the form:

\begin{eqnarray}
H=\left( \matrix{E_{HF}  & 0 & D \cr
                 0  & A  & B \cr
                 D^t  & B^t & C } \right),
\label{matrixH}
\end{eqnarray}

\noindent
where $E_{HF}=\langle HF|H|HF\rangle$ is the Hartree-Fock energy, 
$A_{\sigma'\mu',\sigma\mu}=\langle\sigma'\mu'|H|\sigma\mu\rangle$ is 
the Tamm-Dankoff matrix, $D_{HF,\sigma\rho\mu\lambda}=\langle HF|H|
\sigma\rho,\mu\lambda\rangle$, $B_{\sigma'\mu',\sigma\rho\mu\lambda}=
\langle\sigma'\mu'|H|\sigma\rho,\mu\lambda\rangle$, and
$C_{\sigma'\rho'\mu'\lambda',\sigma\rho\mu\lambda}=
\langle\sigma'\rho',\mu'\lambda'|H|\sigma\rho,\mu\lambda\rangle$. 
$D^t$ and $B^t$ are, respectively, the transposes of matrices $D$ and
$B$. Explicit matrix elements are given in Appendix \ref{appA} for 
completeness.

In sectors with quantum numbers others than the Hartree-Fock state, the
first row and column of matrix (\ref{matrixH}) should be dropped.

An energy cutoff of 3 $\hbar\omega$ in the excitation energy is used 
to control the dimension of the Hamiltonian matrix. The estimated 
error in the ground-state energy is below 0.2 \%. 

\begin{figure}[t]
\begin{center}
\includegraphics[width=.95\linewidth,angle=0]{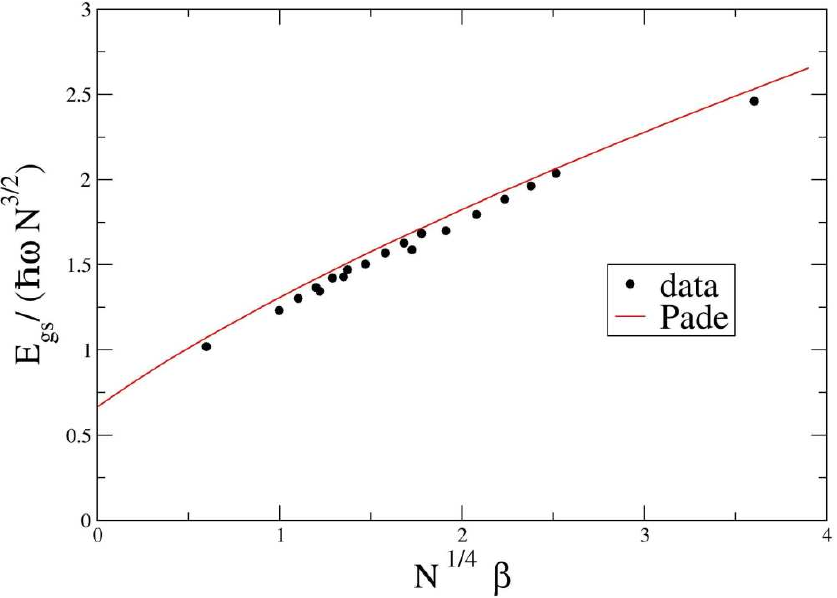}
\caption{\label{fig2} (Color online) Scaling of the ground-state energy 
in medium-sized dots. The large-$N$ expression for the Pad\'e estimate, 
Eq. (\ref{eq5}), is shown as a solid line.}
\end{center}
\end{figure}

We computed the ground-state energy of dots with $N=20$, 30, 42, 56, 72 
and 90, and confinement strengths $\hbar\omega=6$, 12 and 18 meV. 
Notice that these are closed shell quantum dots with ground-state 
angular momentum and spin quantum numbers $L=S=0$. GaAs parameters, 
$m=0.067~m_0$ and $\epsilon=12.8$, were used. We performed
the calculations for three-dimensional dots in which the confinement 
along the symmetry axis (the $z$ axis) is modelled by a rigid-wall well 
of width, $L_z=25$ nm. The constant $N E_z^{(e)}$, where 
$E_z^{(e)}=\hbar^2\pi^2/(2 m L_z^2)$, was removed from the ground-state 
energy. The results are depicted in Fig. \ref{fig2} (dots) along with 
the large-$N$ Pad\'e estimate given by Eq. (\ref{eq5}) (solid line). 
Scaling of the ground-state energy is apparent. The maximum deviations 
with respect to the Pad\'e estimate are below 10 \% for the smallest 
dots with $N=20$. Notice that, for the parameters used in
the calculations, the scaled variable $N^{1/4}\beta$ takes values 
around 1, i.e. in the transition interval from weak to strong coupling
\cite{GPP}. In order to test the whole interval, we use additional 
control cases: one of them deep in the strong coupling regime ($N=42$, 
$\hbar\omega=2$ meV), and the other in the weak coupling
region ($N=20$, $\hbar\omega=50$ meV). They also fit the Pad\'e 
estimate.

\section{Intraband excitations}

\begin{figure}[t]
\begin{center}
\includegraphics[width=.95\linewidth,angle=0]{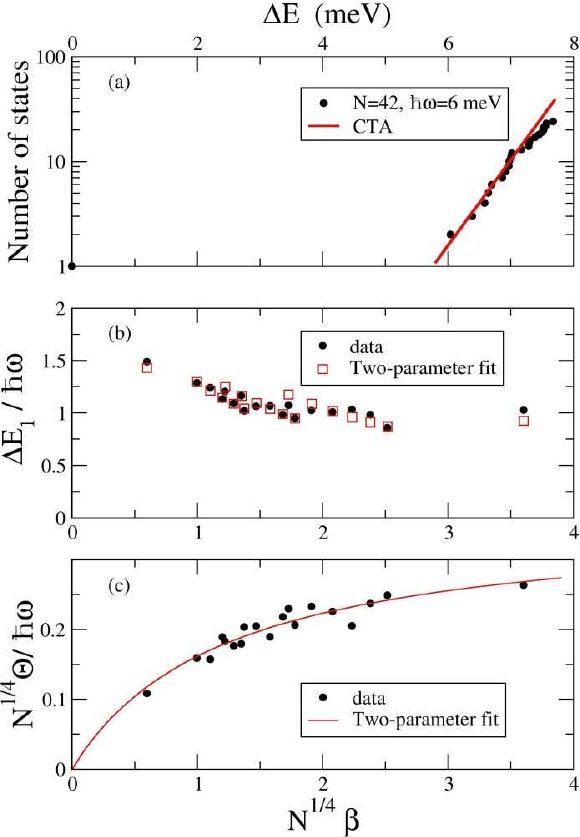}
\caption{\label{fig3} (Color online) (a) The number of excited states 
 in the 42-electron quantum 
 dot as a function of the excitation energy. The confinement strength is 
 $\hbar\omega=6$ meV. (b) The excitation gap to the first excited state as a 
 function of the scaled variable $z=N^{1/4}\beta$. (c) The temperature parameter, 
 $\Theta$, in scaled variables. Fits in (b) and (c) correspond to 
 Eqs. (\ref{eq6}) and (\ref{eq7}).}
\end{center}
\end{figure}

We now turn to the intraband excitations. For simplicity, we consider 
the excited states of the closed-shell quantum dots studied above. We 
restrict the analysis to sectors with the same quantum numbers as the 
ground state, $L=S=0$, in such a way that the ground and excited states 
come out from the same calculation. A sample of the results is shown in 
Fig. \ref{fig3} (a) for the 42-electron dot with confinement 
$\hbar\omega=6$ meV. First, we notice that the excitation gap, which is 
$2~\hbar\omega$  in the noninteracting $\beta\to 0$ limit, is 
renormalized by Coulomb interactions to around 6
meV, that is $1~\hbar\omega$. In the opposite, $\beta\to\infty$, limit, 
the excitation spectrum is that of a big (Wigner) molecule, whose 
small-oscillation frequencies are independent of $\beta$.\cite{GPP} 
The lowest of these frequencies, i.e. that one determining the gap, 
should go to zero for large $N$ in order to meet the acoustic phonon 
of the Wigner lattice. Then, we can look for a simple interpolation 
formula in order to fit the numerical data for the excitation gap:

\begin{equation}
\frac{\Delta E_1}{\hbar\omega} = \frac{2+a_1\beta}
 {1+b_1 N^{\gamma}\beta},
\label{eq6}
\end{equation}

The parameter $\gamma$ appears to be very close to 1/4, thus we fixed 
it to 1/4 and fit again the data in order to obtain $a_1$ and $b_1$. 
The results are shown in Fig. \ref{fig3} (b) as a function of 
$z=N^{1/4}\beta$. We stress that this is only a useful representation 
because $\Delta E_1$ does not scales with $z$, even though Fig. 
\ref{fig3} (b) shows an approximate scaling for intermediate couplings. 
For the parameters $a_1$ and $b_1$, we get: $a_1=3.659$, $b_1=1.878$. 
The result of the fit is excellent, with maximum deviations below 
10 \%, the same as for the ground-state energy. Notice also that 
$N^{1/4} \Delta E_1/(\hbar\omega)$ goes to a universal value, 
$a_1/b_1$, in the strong-coupling limit, $\beta\to\infty$.
Expressions similar to Eq. (\ref{eq6}) for the gap to the first excited 
state should be valid in other angular momentum and spin sectors, and 
also for the energy of collective states (spin- and charge-density 
excitations).

The second point to notice in Fig. \ref{fig3} (a) is the exponential 
growth of the number of states for excitation energies above 6 meV. 
This simple exponential dependence on excitation energy is known in 
Nuclear Physics as the constant temperature approximation (CTA): 
\cite{CTA}

\begin{equation}
N_{states}=N_0 \exp (\Delta E/\Theta).
\label{eq3}
\end{equation}

\noindent
It seems to be a quite general property of the excitation spectrum of
quantum systems. We verified it, for instance, in the energy spectrum of
small quantum dots in strong magnetic fields \cite{Capote}.

We fit the numerical data corresponding to the first 25 excited states 
of the quantum dots mentioned above in order to extract the
``temperature'' parameter, $\Theta$, in Eq. (\ref{eq3}). We took the 
first excited state as the reference of energy. The next 24 states are 
only a few (1 to 4) meV above the first excited state. 

In order to deduce the universal properties of $\Theta$ let us recall 
the $\beta\to 0$ and $\beta\to\infty$ asymptotic regimes. In the 
$\beta\to\infty$ limit, we expect for $\Theta$ a behaviour similar to 
$\Delta E_1$, that is $N^{1/4}\Theta/(\hbar\omega)$ should take a 
universal value. On the other hand, in the $\beta\to 0$ limit the 
excitation energies (with respect to the first excited state) are 
proportional to $\beta$, thus we may write a simple interpolation 
formula for the temperature parameter:

\begin{equation}
\frac{N^{1/4}\Theta}{\hbar\omega}=\frac{a_2 z}{b_2+z},
\label{eq7}
\end{equation}

\noindent
where $a_2=0.360$, $b_2=1.226$. The quality of the fit is also very 
good as can be seen in Fig. \ref{fig3} (c).

\begin{figure}[t]
\begin{center}
\includegraphics[width=.95\linewidth,angle=0]{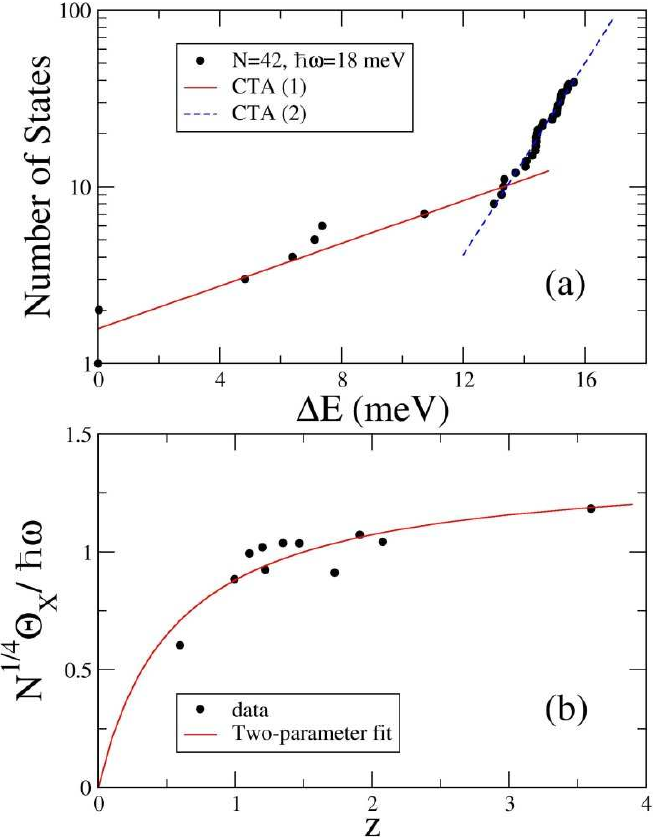}
\caption{\label{fig4} (Color online) (a) The interband (excitonic) 
excitations in the quantum dot with 42 electrons and $\hbar\omega=18$ 
meV. In the $x$ axis the reference energy is the first excitonic state.
(b) Scaling of the temperature parameter for the lowest-energy excitonic 
states.}
\end{center}
\end{figure}

\section{Interband excitations: excitonic states}

Next, we study the interband excitonic excitations of dots with 
$N=20$, 30 and 42, and $\hbar\omega=6$, 12 and 18 meV. The two control 
cases in the strong and weak coupling regimes are also included. 
A basis for excitonic states in these dots is build up in the following 
way: (i) states with one additional electron above the 
Fermi level in the conduction band and a hole in the valence band, 
$|\sigma,\tau\rangle=e_{\sigma}^{\dagger}h_{\tau}^{\dagger}|HF\rangle$ , (ii) states with 
two additional electrons above the Fermi level in the conduction band, 
a hole in the conduction band, and a hole in the valence band, 
$|\sigma\rho,\tau,\mu\rangle=e_{\sigma}^{\dagger}e_{\rho}^{\dagger} 
h_{\tau}^{\dagger}e_{\mu}|HF\rangle$. Details of the computational 
scheme in the present case can be found in Ref.  \onlinecite{ppph}.
The Hartree-Fock single-particle states for holes are 
obtained from the Kohn-Luttinger Hamiltonian in the presence of the 
electronic background. In our model calculations, the oscillator lengths
for electrons, heavy holes and light holes are equal. Kohn-Luttinger 
parameters for GaAs are used \cite{KL}. With a cutoff in the excitation 
energy of 3 $\hbar\omega$, the basis dimension is reduced to around 5000.

We show in Fig. \ref{fig4} (a) a typical spectrum of excitonic 
excitations, corresponding to a dot with $N=42$ and $\hbar\omega=18$ 
meV. The states are characterized by the total angular momentum 
${\cal F}=L_e+L_h-M_h=-3/2$, and total electronic spin projections, 
$S=1/2$. $L_e$ and $L_h$ are orbital angular momenta of electrons and 
holes, respectively, and $M$ is the band momentum of holes along the
$z$ axis. In Fig. \ref{fig4} (a),  
the $x$ axis excitation energies are measured with respect to the first 
excitonic state. The lowest 40 states shown in the figure follow two 
different CTA fits, corresponding to $\Delta E<12$ meV and 
$12<\Delta E<16$ meV. The discontinuity in the slope seems to be a 
quite general fact \cite{Capote} related to different mechanisms of 
formation of the states.

We use the lowest 10 states in order to find a temperature parameter in 
the studied dots, and a law like Eq. (\ref{eq7}) to fit the data. The 
found parameters are: $a_3=1.373$, $b_3=0.559$. It is remarkable that 
the fit performs very good, as shown in Fig. \ref{fig4} (b), signaling 
that the electronic background determines global properties of the 
excitonic excitation spectrum. Unlike the intraband excitations, 
however, we expect the parameters $a_3$ and $b_3$ to depend weakly on 
the dot material (GaAs in this case) because of the Kohn-Luttinger 
Hamiltonian entering the calculations. We shall test in the future to 
what extent this happens.

\begin{figure}[t]
\begin{center}
\includegraphics[width=.9\linewidth,angle=0]{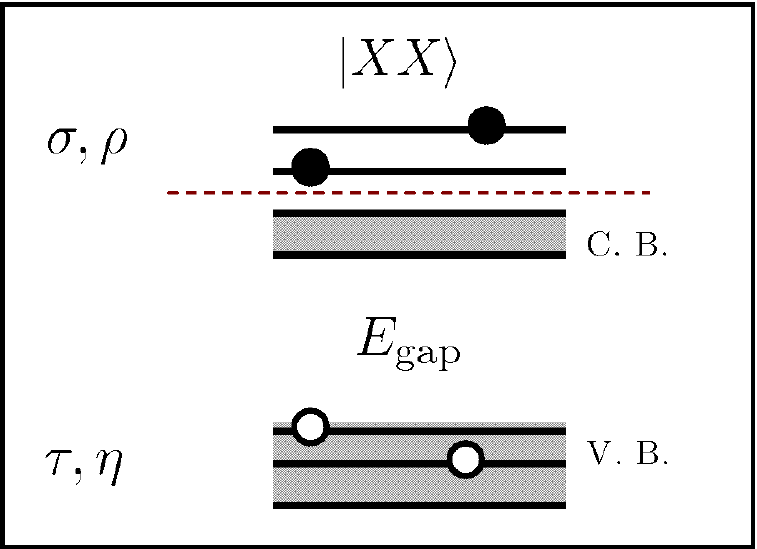}
\caption{\label{fig5} (Color online) Structure of the biexcitonic 
wave function in the Configuration Interaction calculations.}
\end{center}
\end{figure}

\section{Interband excitations: biexcitonic states}

Finally, let us consider the interband biexcitonic excitations in our 
medium-sized dots. The basis functions for the Configuration 
Interaction calculations contains two additional electrons above the 
Fermi level in the conduction band, and two holes in the valence band, 
$|\sigma\rho,\tau\eta\rangle=e_{\sigma}^{\dagger}e_{\rho}^{\dagger}
h_{\tau}^{\dagger}h_{\eta}^{\dagger}|HF\rangle$, with $\sigma < \rho$ 
and $\tau < \eta$. A schematic
representation is given in Fig. \ref{fig5}. In Appendix \ref{appB}, we
give explicit expressions for the Hamiltonian matrix elements.

\begin{figure}[t]
\begin{center}
\includegraphics[width=.95\linewidth,angle=0]{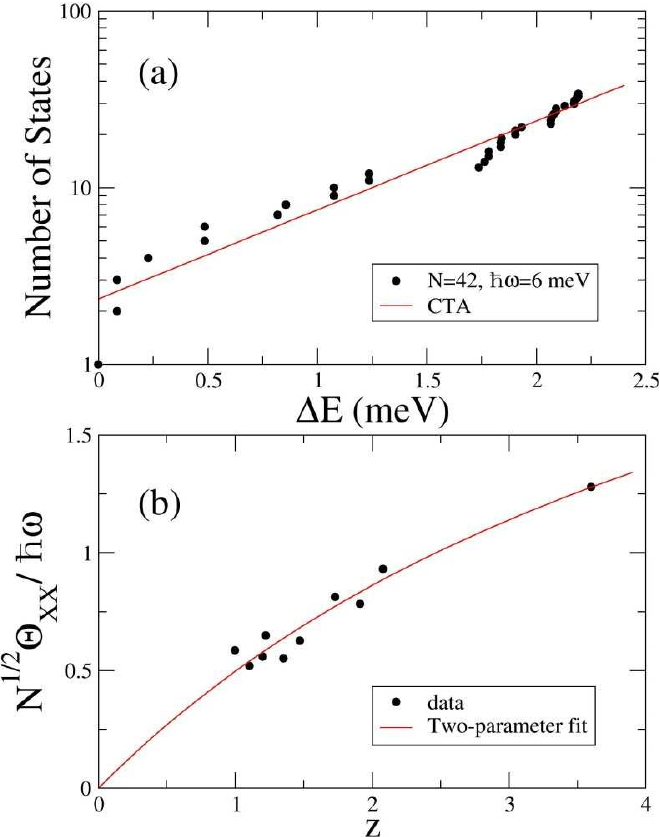}
\caption{\label{fig6} (Color online) (a) The interband (biexcitonic) 
excitations in the quantum dot with 42 electrons and $\hbar\omega=6$ 
meV. In the $x$ axis the reference energy is the first biexcitonic 
state. (b) Scaling of the temperature parameter for the biexcitonic 
excitations.}
\end{center}
\end{figure}

With a cutoff in the excitation energy 
of $2~\hbar\omega$, the Hamiltonian matrix has dimension around 3000.

We draw in Fig. \ref{fig6} (a) the spectrum of biexcitonic excitations 
in the dot with 42 electrons and $\hbar\omega=6$ meV. The quantum 
numbers of the states shown in the figure are: ${\cal F}=0$,  $S=0$. 
We see that in a single CTA fit we may comprise the first 35 states. 
These first states are to be used in the determination of the 
temperature parameter.

The scaling of $\Theta_{xx}$ is shown in the lower panel of Fig. 
\ref{fig6}. Notice the power of $N$, which is now 1/2 instead of 1/4. 
We verified that, by taking this power, the dispersion of points is 
reduced notably. Thus, we fit the data with the formula:

\begin{equation}
\frac{N^{1/2}\Theta_{xx}}{\hbar\omega}=\frac{a_4 z}{b_4+z},
\label{eq8}
\end{equation}

\noindent
where $a_4=3.230$, $b_4=5.503$. The quality of the fit is very good. 
The same comment about the dependence of the parameters on the dot 
material, made above for the excitonic states, applies in the present 
situation.

\section{Conclusions}

In conclusion, we have performed extensive Configuration Interaction 
calculations for medium-sized quantum dots in order to verify universal 
relations for the ground-state energy and the intraband and interband 
(excitonic and biexcitonic) excitation spectrum. The coefficients in 
the r.h.s. of Eqs. (\ref{eq5},\ref{eq6},\ref{eq7}) do not depend
even on the material the dots are made of. On the other hand, 
the coefficients $a_3$, $b_3$, $a_4$, and $b_4$, we believe, are 
specific for GaAs, but independent of $N$ and $\hbar\omega$.

The work can be extended in many directions. We may try to parametrize 
in a universal way the correlation energy, the excitonic and 
biexcitonic binding energies, the excitation gaps to different angular 
momentum and spin sectors, the energy of collective (plasmonic)
excitations, etc. On the other hand, more efforts towards the 
undestanding of the empirical relations obtained for the $\Theta$ 
parameters are needed. Research along these lines is in progress.

\begin{acknowledgments}
Part of this work was performed using the computing facilities of the 
Abdus Salam ICTP, Trieste, Italy. The authors acknowledge support by 
the Caribbean Network for Quantum Mechanics, Particles and Fields (ICTP) 
and by the Programa Nacional de Ciencias B\'asicas (Cuba).
\end{acknowledgments}

\appendix

\section{Explicit matrix elements for intraband excitations}
\label{appA}

In Eq. (\ref{matrixH}), $E_{HF}$ is the Hartree-Fock total energy
\cite{Ring}:

\begin{equation}
E_{HF}=\frac{1}{2}\sum_{\mu\le\mu_F}\left\{\varepsilon^{(e)}_{\mu}
 +\sum_{k,l,S_z}|R^{(\mu)}_{klS_z}|^2 \varepsilon^{(0)}_{kls_z}
 \right\},
\end{equation}

\noindent
where $\mu_F$ is the Fermi level, $\varepsilon^{(e)}_{\mu}$ is the
Hartree-Fock energy of the electron state $\mu$, 
$\varepsilon^{(0)}_{klS_z}$ is the energy of 2D oscillator states, 
characterized by the quantum numbers $k$ (radial number), $l$ 
(angular momentum), and $S_z$ (spin projection). That is:

\begin{equation}
\varepsilon^{(0)}_{kls_z}=E_z^{(e)}+\hbar\omega(2 k+|l|+1).
\end{equation}

\noindent
The state $\mu$ is expanded in oscillator states as follows:

\begin{equation}
|\mu\rangle=\sum_{k,l,S_z} R^{(\mu)}_{k,l,S_z} |k,l,S_z\rangle.
\label{expansion}
\end{equation}

\noindent
In the studied closed-shell dots, $l$ and $S_z$ are good quantum numbers
of $|\mu\rangle$, and the above sum run only over $k$.

On the other hand, in Eq. (\ref{matrixH}) $A$ is the Tamm-Dankoff 
matrix\cite{Ring}: 

\begin{eqnarray}
A_{\sigma'\mu',\sigma\mu}&=&\left(E_{HF}+\varepsilon^{(e)}_{\sigma}-
 \varepsilon^{(e)}_{\mu}\right)\delta_{\sigma\sigma'}\delta_{\mu\mu'}
 \nonumber\\
  &+&\beta \langle\sigma',\mu|1/r_{ee}|\widetilde{\mu',\sigma}\rangle,
\end{eqnarray}

\noindent
where the antisymmetrized Coulomb matrix elements are defined as:

\begin{equation}
\langle\sigma',\mu|1/r_{ee}|\widetilde{\mu',\sigma}\rangle=
\langle\sigma',\mu|1/r_{ee}|\mu',\sigma\rangle-
\langle\sigma',\mu|1/r_{ee}|\sigma,\mu'\rangle.
\end{equation}

Coulomb matrix elements $\langle\sigma',\mu|1/r_{ee}|\mu',\sigma\rangle$
are computed in terms of matrix elements among oscillator states by 
using the expansion (\ref{expansion}). 

Finally, matrices $D$, $B$ and $C$ are explicitly written as:

\begin{equation}
D_{HF,\sigma\rho\mu\lambda}=\beta\langle\mu,\lambda|1/r_{ee}|
 \widetilde{\rho,\sigma}\rangle.
\end{equation}

\begin{eqnarray}
B_{\sigma'\mu',\sigma\rho\mu\lambda}&=&\beta\left\{
 \langle\mu,\lambda|1/r_{ee}|\widetilde{\mu',\rho}\rangle
 \delta_{\sigma\sigma'} \right.\nonumber\\
 &+&\langle\mu,\lambda|1/r_{ee}|\widetilde{\sigma,\mu'}\rangle
 \delta_{\rho\sigma'} \nonumber\\
 &+&\langle\sigma',\lambda|1/r_{ee}|\widetilde{\rho,\sigma}\rangle
 \delta_{\mu\mu'} \nonumber\\ 
 &+&\left.\langle\sigma',\mu|1/r_{ee}|\widetilde{\sigma,\rho}\rangle
 \delta_{\lambda\mu'}\right\}.
\end{eqnarray}

\begin{eqnarray}
&&C_{\sigma'\rho'\mu'\lambda',\sigma\rho\mu\lambda}=\nonumber\\
 &&\left ( E_{HF}+\varepsilon^{(e)}_{\sigma}+\varepsilon^{(e)}_{\rho}
 -\varepsilon^{(e)}_{\mu}-\varepsilon^{(e)}_{\lambda}\right )
 \delta_{\sigma\sigma'}\delta_{\rho\rho'}\delta_{\mu\mu'}
 \delta_{\lambda\lambda'}\nonumber\\
 &&+\beta\bigg\lbrace
 \langle\mu,\lambda|1/r_{ee}|\widetilde{\mu',\lambda'}\rangle
 \delta_{\sigma\sigma'}\delta_{\rho\rho'}\nonumber\\
 &&+\langle\rho',\lambda|1/r_{ee}|\widetilde{\lambda',\rho}\rangle
 \delta_{\sigma\sigma'}\delta_{\mu\mu'}
 +\langle\rho',\mu|1/r_{ee}|\widetilde{\rho,\lambda'}\rangle
 \delta_{\sigma\sigma'}\delta_{\lambda\mu'}\nonumber\\
 &&+\langle\rho',\lambda|1/r_{ee}|\widetilde{\rho,\mu'}\rangle
 \delta_{\sigma\sigma'}\delta_{\mu\lambda'}
 +\langle\rho',\mu|1/r_{ee}|\widetilde{\mu',\rho}\rangle
 \delta_{\sigma\sigma'}\delta_{\lambda\lambda'}\nonumber\\
 &&+\langle\rho',\lambda|1/r_{ee}|\widetilde{\sigma,\lambda'}\rangle
 \delta_{\mu\mu'}\delta_{\rho\sigma'}
 +\langle\sigma',\lambda|1/r_{ee}|\widetilde{\rho,\lambda'}\rangle
 \delta_{\mu\mu'}\delta_{\sigma\rho'}\nonumber\\
 &&+\langle\sigma',\lambda|1/r_{ee}|\widetilde{\lambda',\sigma}\rangle
 \delta_{\mu\mu'}\delta_{\rho\rho'}
 +\langle\rho',\sigma'|1/r_{ee}|\widetilde{\rho,\sigma}\rangle
 \delta_{\mu\mu'}\delta_{\lambda\lambda'}\nonumber\\
 &&+\langle\rho',\mu|1/r_{ee}|\widetilde{\lambda',\sigma}\rangle
 \delta_{\lambda\mu'}\delta_{\rho\sigma'}
 +\langle\sigma',\mu|1/r_{ee}|\widetilde{\lambda',\rho}\rangle
 \delta_{\lambda\mu'}\delta_{\sigma\rho'}\nonumber\\
 &&+\langle\sigma',\mu|1/r_{ee}|\widetilde{\sigma,\lambda'}\rangle
 \delta_{\lambda\mu'}\delta_{\rho\rho'}
 +\langle\rho',\lambda|1/r_{ee}|\widetilde{\mu',\sigma}\rangle
 \delta_{\mu\lambda'}\delta_{\rho\sigma'}\nonumber\\
 &&+\langle\sigma',\lambda|1/r_{ee}|\widetilde{\sigma,\mu'}\rangle
 \delta_{\mu\lambda'}\delta_{\rho\rho'}
 +\langle\sigma',\lambda|1/r_{ee}|\widetilde{\mu',\rho}\rangle
 \delta_{\mu,\lambda'}\delta_{\sigma\rho'}\nonumber\\
 &&+\langle\rho',\mu|1/r_{ee}|\widetilde{\sigma,\mu'}\rangle
 \delta_{\lambda\lambda'}\delta_{\rho\sigma'}
 +\langle\sigma',\mu|1/r_{ee}|\widetilde{\rho,\mu'}\rangle
 \delta_{\lambda\lambda'}\delta_{\sigma\rho'}\nonumber\\
 &&+\langle\sigma',\mu|1/r_{ee}|\widetilde{\mu',\sigma}\rangle
 \delta_{\lambda\lambda'}\delta_{\rho\rho'}\bigg\rbrace.
\end{eqnarray}

\section{Explicit matrix elements for biexcitonic excitations}
\label{appB}

In the biexcitonic sector, the Hamiltonian matrix elements take the
form:

\begin{eqnarray}
&&\langle\sigma'\rho',\tau'\eta'|H|\sigma\rho,\tau\eta\rangle=\nonumber\\
 &&\left(E_{HF}+\varepsilon^{(e)}_\sigma+\varepsilon^{(e)}_\rho+
   \varepsilon^{(h)}_\tau+\varepsilon^{(h)}_\eta\right)
   \delta_{\sigma\sigma'}\delta_{\rho\rho'}
   \delta_{\tau\tau'}\delta_{\eta\eta'}
 \nonumber \\
 &&+\beta \langle\sigma',\rho'|1/r_{ee}|\widetilde{\sigma,\rho}\rangle
  \delta_{\tau\tau'}\delta_{\eta\eta'}\nonumber\\
 &&+\beta \langle\tau',\eta'|1/r_{hh}|\widetilde{\tau,\eta}\rangle
  \delta_{\sigma\sigma'}\delta_{\rho\rho'} \nonumber \\ 
 &&-\beta\bigg\lbrace \langle\rho',\eta'|1/r_{eh}|\rho,\eta\rangle
  \delta_{\tau\tau'}\delta_{\sigma\sigma'}\nonumber\\
 &&-\langle\rho',\eta'|1/r_{eh}|\rho,\tau\rangle
  \delta_{\eta\tau'}\delta_{\sigma\sigma'} \nonumber \\
 &&-\langle\rho',\tau'|1/r_{eh}|\rho,\eta\rangle
  \delta_{\tau\eta'}\delta_{\sigma\sigma'}+
  \langle\rho',\tau'|1/r_{eh}|\rho,\tau\rangle
  \delta_{\eta\eta'}\delta_{\sigma\sigma'} \nonumber \\
 &&-\langle\rho',\eta'|1/r_{eh}|\sigma,\eta\rangle
  \delta_{\tau\tau'}\delta_{\rho\sigma'}+
  \langle\rho',\eta'|1/r_{eh}|\sigma,\tau\rangle
  \delta_{\eta\tau'}\delta_{\rho\sigma'} \nonumber \\  
 &&+\langle\rho',\tau'|1/r_{eh}|\sigma,\eta\rangle
  \delta_{\tau\eta'}\delta_{\rho\sigma'}-
  \langle\rho',\tau'|1/r_{eh}|\sigma,\tau\rangle
  \delta_{\eta\eta'}\delta_{\rho\sigma'} \nonumber \\  
 &&-\langle\sigma',\eta'|1/r_{eh}|\rho,\eta\rangle
  \delta_{\tau\tau'}\delta_{\sigma\rho'}+
  \langle\sigma',\eta'|1/r_{eh}|\rho,\tau\rangle
  \delta_{\eta\tau'}\delta_{\sigma\rho'} \nonumber \\  
 &&+\langle\sigma',\tau'|1/r_{eh}|\rho,\eta\rangle
  \delta_{\tau\eta'}\delta_{\sigma\rho'}-
  \langle\sigma',\tau'|1/r_{eh}|\rho,\tau\rangle
  \delta_{\eta\eta'}\delta_{\sigma\rho'} \nonumber \\  
 &&+\langle\sigma',\eta'|1/r_{eh}|\sigma,\eta\rangle
  \delta_{\tau\tau'}\delta_{\rho\rho'}-
  \langle\sigma',\eta'|1/r_{eh}|\sigma,\tau\rangle
  \delta_{\eta\tau'}\delta_{\rho\rho'} \nonumber \\  
 &&-\langle\sigma',\tau'|1/r_{eh}|\sigma,\eta\rangle
  \delta_{\tau\eta'}\delta_{\rho\rho'}+ 
  \langle\sigma',\tau'|1/r_{eh}|\sigma,\tau\rangle
  \delta_{\eta\eta'}\delta_{\rho\rho'} \nonumber \\  
 &&-\sum_{\alpha\le\mu_F}\big[
  \langle\alpha,\eta'|1/r_{eh}|\alpha,\eta\rangle
  \delta_{\tau\tau'}-
  \langle\alpha,\eta'|1/r_{eh}|\alpha,\tau\rangle
  \delta_{\eta\tau'} \nonumber \\  
 &&-\langle\alpha,\tau'|1/r_{eh}|\alpha,\eta\rangle
  \delta_{\tau\eta'}\nonumber\\
 &&+\langle\alpha,\tau'|1/r_{eh}|\alpha,\tau\rangle
  \delta_{\eta\eta'}\big]
  \delta_{\sigma\sigma'}\delta_{\rho\rho'} \bigg\rbrace.
\label{xx12}
\end{eqnarray}

The total Hamiltonian, $H$, in addition to the terms of Eq. (\ref{eq1}),
now includes the single-particle energy of holes, electron-hole, and
hole-hole interactions. The Hartree-Fock electron and hole states 
should be expanded in oscillator functions when Coulomb matrix elements 
are to be computed. For hole states, coming from a Kohn-Luttinger
Hartree-Fock problem, we have the expansion:

\begin{equation}
|\tau\rangle=\sum_{k,l,m,k_z} R^{(\tau)}_{k,l,m,k_z} |k,l,m,k_z\rangle,
\label{expansionh}
\end{equation}

\noindent
where $k$ and $l$ are oscillator quantum numbers, $m=\pm 3/2,\pm 1/2$
is hole (band) angular momentum projection, and $k_z=1,\dots, 6$ 
labels sub-band states in the well. The relatively large number of terms
entering the expansion (\ref{expansionh}) makes the calculation of
Coulomb matrix elements for holes lengthy.


\begin{thebibliography}{99}
\bibitem{TF} L.H. Thomas, Proc. Camb. Phil. Soc. 23, 542 (1927);
 E. Fermi, Rend. Accad. Naz. Lincei 6, 602 (1927).
\bibitem{Kirzhnits} D.A. Kirzhnits, Yu.E. Lozovik, and G.V.
 Shpatakovskaya, Sov. Phys. Usp. 18, 649 (1975).
\bibitem{Lieb1} E.H. Lieb, Rev. Mod. Phys. 53, 603 (1981).
\bibitem{Spruch} L. Spruch, Rev. Mod. Phys. 63, 1512 (1991).
\bibitem{KM} L. Kouwenhoven and C. Marcus, Phys. World 11 (6), 35 
 (1998).
\bibitem{Hawrylak}  L. Jacak, P. Hawrylak, and A. Wojs, {\it Quantum dots} 
 (Springer-Verlag, Berlin, 1998). 
\bibitem{Serra} Ll. Serra and A. Puente, Eur. Phys. J. D 14, 77 (2001).
\bibitem{Ullmo} D. Ullmo, H. Jiang, W. Yang, and H.U. Baranger,
 Phys. Rev. B 70, 205309 (2004). 
\bibitem{Lieb2} E.H. Lieb, J.P. Solovej, and J. Yngvason, Phys. Rev. 
 B 51, 10646 (1995).
\bibitem{TFmio} A.Gonzalez and R. Gonzalez, Sov. Phys. - Lebedev Inst.
 Reports, 3, 36 (1990). 
\bibitem{PLA} A. Belov, Yu.E. Lozovik, and A. Gonzalez, Phys. Lett. A
 142, 389 (1989).
\bibitem{GPP} A. Gonzalez, B. Partoens and F.M. Peeters, Phys. Rev. 
 B 56, 15740 (1997). 
\bibitem{CI} C.J. Cramer, {\it Essentials of Computational Chemistry:
 Theories and Models} (Wiley, Chichester, 2006).
\bibitem{Ring} P. Ring and P. Schuck, {\it The Nuclear Many-Body Problem}
 (Springer-Verlag, New-York, 1980).  
\bibitem{CTA} T. Ericson, Adv. Phys. 9, 425 (1960); A. Gilbert,
 A.G.W. Cameron, Can. J. Phys. 43, 1446 (1965). 
\bibitem{Capote} A. Gonzalez and R. Capote, Phys. Rev. B 66, 113311 
 (2002).
\bibitem{ppph} A. Odriazola, A. Delgado, and A. Gonzalez, Phys. Rev. 
 B 78, 035329 (2008).
\bibitem{KL} Landolt-Bornstein, {\it Numerical Data and Functional 
 Relationship in Science and Technology}, Group III, Volume 17 
 (Springer-Verlag, Berlin, 1982).  
\end{thebibliography}
\end{document}